\documentclass[amsmath,amssymb,amsbsy,reprint,prb,preprintnumbers,showpacs,superscriptaddress]{revtex4-2}
\usepackage{graphicx,color}
\usepackage{dcolumn}
\usepackage{bm}
\usepackage{braket}
\usepackage{mathtools}
\usepackage{ulem}
\usepackage[breaklinks,colorlinks=true,linkcolor=blue,urlcolor=blue,citecolor=blue]{hyperref}
\usepackage{times}
\usepackage{physics}
\usepackage{latexsym}
\usepackage{amsmath, amssymb}
\usepackage{mathtools}
\usepackage{multirow}

\newcommand{\be}{\begin{eqnarray}}
\newcommand{\ee}{\end{eqnarray}}

\begin{document}

\title{Fate of average symmetry-protected topological states under symmetry-preserving quantum operations}
\date{\today}
\author{Yoshihito Kuno} 
\affiliation{Graduate School of Engineering Science, Akita University, Akita 010-8502, Japan}
\author{Takahiro Orito}
\affiliation{Department of Physics, College of Humanities and Sciences, Nihon University, Sakurajosui, Setagaya, Tokyo 156-8550, Japan}


\begin{abstract}
The fate of average symmetry-protected topological (ASPT) states based on the one-dimensional cluster state is investigated by using the Choi mapping, namely, the doubled Hilbert-space formalism. By introducing two effective spins, $\tau$ and $\eta$, on each doubled rung, we obtain a transparent representation of the ASPT parent Hamiltonian and systematically identify the conserved stabilizers and the corresponding orders. 
We further study symmetry-preserving operations that deform the ASPT by combining qualitative effective-Hamiltonian and systematic parent-Hamiltonian analyses with numerical matrix-product-state filtering calculations. 
We find that although the ASPT remains robust over a broad range of decoherence strengths, strong-to-weak spontaneous symmetry breaking (SWSSB) becomes exact in the maximal-decoherence limit. In addition, we consider symmetry-preserving no-click postselection, which removes the exact conservation law supporting the R\'enyi-2 string order of the ASPT and drives the system toward weak-symmetry spontaneous symmetry breaking (WSSB). Again, the ASPT remains robust over a broad range of postselection strengths, with WSSB becoming pronounced only near the projective limit and exact at the limit itself.
\end{abstract}


\maketitle
\section{Introduction} 
The characterization of quantum phases in open systems has recently attracted considerable attention, extending the notion of symmetry-protected topological (SPT) phases \cite{Pollman2010,Pollman2012,Chen2013,Chen2014,Senthil2015} beyond pure ground states. In particular, average symmetry-protected topological (ASPT) phases \cite{ma2023,ma2024,Lee2025,Ma_PRXQuantum6,xue2024tensornetwork} provide a useful framework for describing mixed states in which decoherence reduces a strong symmetry to a weak symmetry \cite{Buca2012,groot2022}. 
Such states can retain nonlocal information inherited from the parent SPT state, even though conventional pure-state diagnostics are no longer directly applicable. An important question is therefore how robust ASPT order is against further symmetry-preserving quantum operations and, when it eventually disappears, what type of order replaces it. Previous studies suggest that one-dimensional ASPT states can be robust against certain symmetry-preserving quantum operations \cite{bao2023,Seunghun2025,Lee2025,Shah2025instability}. 
However, a systematic understanding of this robustness and of the competing orders that emerge as ASPT order is suppressed remains lacking.

In this work, we address this question for the one-dimensional cluster SPT state \cite{Briegel2009,Son2011,Son2012} by employing the Choi mapping \cite{JAMIOLKOWSKI1972,Choi1975}, which represents a mixed state as a pure state in a doubled Hilbert space. 
It has recently been shown that the one-dimensional cluster SPT state can be converted into an ASPT state by an even-site $Z$-decoherence channel \cite{Lee2025,Ma_PRXQuantum6}. 
We investigate the properties of the ASPT by introducing two effective spins, $\tau$ and $\eta$, on each doubled rung. In this representation, the ASPT parent Hamiltonian takes a transparent form in terms of the cluster operators $C^\tau_j$ and $C^\eta_j$. This allows us to identify which local stabilizers remain exactly conserved after decoherence and to establish a direct correspondence between these stabilizers and several conventional and R\'enyi-2 string orders \cite{Lee2025,Ma_PRXQuantum6}. In particular, the $\tau$-$\eta$ representation provides a unified description of the different string orders previously used to characterize ASPT states. 
In addition, related ASPT-like structures have been discussed for higher-dimensional cluster states, where higher-form symmetries \cite{Guo-and-Ashida2024} and their associated anomalies play an essential role \cite{Lee2025}. 
Here, we restrict ourselves to the one-dimensional cluster state, for which the doubled-space structure permits a particularly transparent microscopic spin mapping and a detailed analysis of the competing orders under symmetry-preserving operations.

We then investigate the fate of the ASPT under two distinct quantum operations that preserve both its strong and weak symmetries. First, we introduce an additional even-site $X$-decoherence channel. The qualitative effective Hamiltonian contains a transverse-field Ising model (TFIM) \cite{PFEUTY197079} sector, suggesting a competition between the ASPT string order and the R\'enyi-2 $ZZ$ long-range order associated with strong-to-weak spontaneous symmetry breaking (SWSSB). The exact parent-Hamiltonian construction confirms that the transverse-field-like perturbation appears at leading order in the weak-filtering expansion, while higher-order terms generate additional correlated interactions. In the maximal-decoherence limit, the doubled-space representation of the channel becomes a projector, and the surviving stabilizer constraints directly enforce exact $ZZ$ long-range order. Thus, both the weak-filtering regime and the maximal-decoherence limit admit analytical descriptions.

Second, we consider a no-click postselection protocol generated by weak measurements. In the doubled space, the postselection acts as a nonunitary filter coupling the $\tau$ and $\eta$ sectors. A mean-field treatment of the qualitative effective Hamiltonian suggests an Ising-like instability toward a WSSB state associated with the diagonal $Z_2$ symmetry. However, an expansion of the exact parent Hamiltonian shows that additional correlated interactions appear at the same leading order as the perturbation retained in the mean-field description. Therefore, the Ising-like instability of the qualitative effective Hamiltonian does not by itself establish a phase transition at finite postselection strength in the actual postselected state. In the projective limit, by contrast, the projector constraints directly enforce exact WSSB order.

Finally, we use matrix-product-state filtering calculations to determine how these analytically controlled regimes are connected at finite decoherence and postselection strengths. Our results provide a systematic picture of the fate and robustness of one-dimensional ASPT order, showing how ASPT string order is progressively suppressed while symmetry-breaking correlations are enhanced under these operations and become exact only in the corresponding limiting cases.

The rest of this paper is organized as follows. 
In Sec.~II, we introduce the cluster model, the even-site $Z$-decoherence channel \cite{Nielsen2011}, and the resulting ASPT state. 
In Sec.~III, we introduce the $\tau$-$\eta$ spin mapping in the doubled Hilbert space, which identifies the conserved quantities and the corresponding correlators, thereby clarifying the order structure of the ASPT. This constitutes the first main result of this work.
In Sec.~IV, using the $\tau$-$\eta$ spin mapping, we examine the competition between ASPT and SWSSB orders under the even-site $X$-decoherence channel through qualitative effective-Hamiltonian and exact parent-Hamiltonian analyses, including the maximal-decoherence limit.
In Sec.~V, we examine the competition between ASPT and WSSB orders under symmetry-preserving no-click postselection through qualitative effective-Hamiltonian and exact parent-Hamiltonian analyses, including the projective limit.
In Sec.~VI, we use density-matrix renormalization group and matrix-product-state filtering calculations to investigate both the even-site $X$-decoherence channel and no-click postselection. We show that the ASPT remains robust over a broad range of finite strengths, while the symmetry-breaking correlations become pronounced near the maximal-decoherence and projective limits, respectively, through strong finite-size crossovers.
Section~VII is devoted to the discussion and conclusion.

\section{ASPT based on the cluster SPT state}
We begin with the one-dimensional cluster model \cite{Suzuki1971,Bartlett2009,Smacchia2011,Giampaolo2015,Lahtinen2015}:
\begin{eqnarray}
H_{\rm c}=-\sum^{L-1}_{j=0}K_j,
\end{eqnarray}
where $K_j\equiv Z_{j-1}X_jZ_{j+1}$ is a stabilizer, $L$ is an even system size, and periodic boundary conditions are imposed. The model has a $Z_2\times Z_2$ symmetry generated by the even- and odd-sublattice spin-flip operators $W\equiv\prod_{j\in{\rm even}}X_j$ and $S\equiv\prod_{j\in{\rm odd}}X_j$, respectively \cite{Son2012}. 
The Hamiltonian $H_{\rm c}$ has a unique gapped ground state, which is the cluster SPT state \cite{Son2011,Son2012}. The symmetries generated by $W$ and $S$ are both strong symmetries of this pure state (see Appendix B).  

Throughout this work, we represent mixed states and quantum channels using the Choi mapping and its associated doubled-Hilbert-space notation \cite{Choi1975,JAMIOLKOWSKI1972}. 
Under this mapping, a density matrix $\rho$ acting on $\mathcal{H}$ is represented as a vector in the doubled Hilbert space $\mathcal{H}_{u}\otimes\mathcal{H}_{\ell}$, where $u$ and $\ell$ denote the upper and lower Hilbert spaces associated with the bra and ket indices, respectively. The Choi mapping is briefly reviewed in Appendix A.

Following Ref.~\cite{Lee2025}, an ASPT state can be generated by the even-site $Z$-decoherence channel, whose doubled-space representation is
\begin{eqnarray}
\hat{\mathcal{E}}_{Z,e}&=&\prod^{L/2-1}_{m=0}\hat{\mathcal{E}}_{Z,2m},\nonumber\\
\hat{\mathcal{E}}_{Z,2m}&=&\biggl[(1-p_z)\hat{I}^*_{2m,u} \otimes \hat{I}_{2m,\ell}+p_z Z^*_{2m,u}\otimes Z_{2m,\ell}\biggr],\nonumber\\
\end{eqnarray}
where $p_z$ is the decoherence strength, with $0\leq p_z\leq 1/2$. At $p_z=1/2$, the channel corresponds to a nonselective projective measurement in the $Z$ basis. For $p_z>0$, it converts the pure cluster SPT state into a mixed state and reduces the strong symmetry generated by $W$ to a weak symmetry. Although $\hat{\mathcal{E}}_{Z,e}$ is trace preserving as a quantum channel, its doubled-space representation $\hat{\mathcal{E}}_{Z,e}$ is nonunitary and therefore does not preserve the Hilbert-space norm of the Choi vector.
We focus on the doubled state representing the resulting mixed state,
$|\rho^D\rangle\rangle\equiv\hat{\mathcal{E}}_{Z,e}|\rho_0\rangle\rangle$, where $\rho_0=|\psi_0\rangle\langle\psi_0|$, $|\rho_0\rangle\rangle\equiv |\psi^*_0\rangle_u\otimes|\psi_0\rangle_\ell$, and $|\psi_0\rangle$ is the cluster-SPT ground state of $H_{\rm c}$ \cite{lee2023,Grover2024,Lee2025}. The state $|\rho^D\rangle\rangle$ represents the ASPT state considered below.

Following Ref.~\cite{Lee2025}, the ASPT parent Hamiltonian is defined in the doubled Hilbert space with a ladder geometry, and its ground state is the ASPT state $|\rho^D\rangle\rangle$. The Hamiltonian is given by 
\begin{eqnarray}
H_{\rm eff}&=&H_{{\rm c},u}+H_{{\rm c},\ell}+H_{\rm rung}+({\rm const.}).\label{Heff}\\
H_{{\rm c},u}
&=&
\sum^{L/2-1}_{m=0} \biggl[-\frac{\cosh 2\beta}{2}Z_{2m-1,u}X_{2m,u}Z_{2m+1,u}\nonumber\\
&&-\frac{1}{2}Z_{2m,u}X_{2m+1,u}Z_{2m+2,u}\biggr].\\
H_{{\rm c},\ell}
&=&
\sum^{L/2-1}_{m=0} \biggl[-\frac{\cosh 2\beta}{2}Z_{2m-1,\ell}X_{2m,\ell}Z_{2m+1,\ell}\nonumber\\
&&-\frac{1}{2}Z_{2m,\ell}X_{2m+1,\ell}Z_{2m+2,\ell}\biggr].\\
H_{\rm rung}
&=&
-\frac{\sinh 4\beta}{2}\sum^{L/2-1}_{m=0} Z_{2m,u}Z_{2m,\ell},
\end{eqnarray}
where $\beta=-\frac12\ln(1-2p_z)$.
The derivation is shown in Appendix C. 

The strong and weak symmetries protecting the ASPT are explicit in the form of $H_{\rm eff}$. The odd-sublattice symmetry generated by $S$ remains strong and acts independently on the upper and lower copies:
\begin{eqnarray}
(S^*_{u}\otimes I_{\ell})H_{\rm eff}(S^*_{u}\otimes I_{\ell})^\dagger=H_{\rm eff},\label{strong_sym1}\\
(I_{u}\otimes S_{\ell})H_{\rm eff}(I_{u}\otimes S_{\ell})^\dagger=H_{\rm eff}.\label{strong_sym2}
\end{eqnarray}

By contrast, the even-sublattice symmetry generated by $W$ survives only as a diagonal weak symmetry:
\begin{eqnarray}
(W^*_{u}\otimes W_{\ell})H_{\rm eff}(W^*_{u}\otimes W_{\ell})^\dagger=H_{\rm eff}.\label{weak_sym}
\end{eqnarray}
For $p_z>0$, the rung term $H_{\rm rung}$ breaks the independent actions of $W$ on the upper and lower copies, leaving only the diagonal weak symmetry \cite{Lee2025}.

The ASPT state $|\rho^D\rangle\rangle$ is protected by these symmetries, consistently with the classification framework of Ref.~\cite{Ma_PRXQuantum6}.

\section{$\tau$-$\eta$ spin mapping for the ASPT}
We now introduce the $\tau$-$\eta$ spin mapping, which makes the doubled-space structure of the ASPT transparent. The two physical spins on each rung span a four-dimensional local Hilbert space, which we represent using two effective spin-$1/2$ degrees of freedom, $\tau$ and $\eta$, defined as follows \cite{ziereis2025strongtoweaksymmetrybreakingphases}.
For $j=2m+1$, 
\begin{align}
\tau_j^x &= X_{j,u}X_{j,\ell},
&
\tau_j^z &= Z_{j,u},
\nonumber\\
\eta_j^x &= X_{j,\ell},
&
\eta_j^z &= Z_{j,u}Z_{j,\ell}.
\end{align}
For $j=2m$, we define
\begin{align}
\tau_j^x &= X_{j,u},
&
\tau_j^z &= Z_{j,u}Z_{j,\ell},
\nonumber\\
\eta_j^x &= X_{j,u}X_{j,\ell},
&
\eta_j^z &= Z_{j,\ell}.
\end{align}
These transformations preserve the Pauli algebra.

Applying this mapping to the ASPT parent Hamiltonian gives
\begin{align}
H_{\mathrm{eff}}
=
&-\frac{\cosh 2\beta}{2}
\sum_m
C_{2m}^{\tau}
\left(1+C_{2m}^{\eta}\right)
\nonumber\\
&-\frac{1}{2}
\sum_m
C_{2m+1}^{\eta}
\left(1+C_{2m+1}^{\tau}\right)
\nonumber\\
&-\frac{\sinh 4\beta}{2}
\sum_m \tau_{2m}^{z}+({\rm const.}),
\end{align}
where we define the cluster operators
\begin{eqnarray}
C_j^{\tau}
=\tau_{j-1}^{z}\tau_j^{x}\tau_{j+1}^{z}
=
\begin{cases}
K_{j,u}, & j\ {\rm even}\\
K_{j,u}K_{j,\ell}, & j\ {\rm odd},
\end{cases}
\end{eqnarray}
and
\begin{eqnarray}
C_j^{\eta}
=
\eta_{j-1}^{z}\eta_j^{x}\eta_{j+1}^{z}
=
\begin{cases}
K_{j,u}K_{j,\ell}, & j\ {\rm even}\\
K_{j,\ell}, & j\ {\rm odd}
\end{cases}.
\end{eqnarray}
Here $K_{j,a}\equiv Z_{j-1,a}X_{j,a}Z_{j+1,a}$ with $a=u,\ell$. 

The cluster operators mutually commute:
$[C_i^{\tau},C_j^{\tau}]=0$, 
$[C_i^{\eta},C_j^{\eta}]=0$, and
$[C_i^{\tau},C_j^{\eta}]=0$.
They further satisfy
$[C_{2m+1}^{\tau},H_{\rm eff}]=0$
and
$[C_j^{\eta},H_{\rm eff}]=0$
for any $j$, whereas
$[C_{2m}^{\tau},H_{\rm eff}]\neq 0$ for $\beta\neq0$.
Thus, $C^\tau_{2m+1}$ and $C^\eta_j$ are exact local conserved quantities, whereas $C^\tau_{2m}$ is not conserved and remains dynamical.
In the state $|\rho^D\rangle\rangle$, the conserved quantities have eigenvalue $+1$. In particular,
$C^{\tau}_{2m+1}|\rho^D\rangle\rangle
=|\rho^D\rangle\rangle$
since
$[C^{\tau}_{2m+1},\hat{\mathcal{E}}_{Z,e}]=0$
and
$C^{\tau}_{2m+1}|\rho_{0}\rangle\rangle
=|\rho_{0}\rangle\rangle$.
Similarly,
$[C^{\eta}_{j},\hat{\mathcal{E}}_{Z,e}]=0$
and
$C^{\eta}_{j}|\rho_{0}\rangle\rangle
=|\rho_{0}\rangle\rangle$
for any $j$, and hence
$C^{\eta}_{j}|\rho^D\rangle\rangle
=|\rho^D\rangle\rangle$.
Therefore, the ASPT ground state of $H_{\rm eff}$ lies in the sectors
$C^{\tau}_{2m+1}=+1$ and $C^{\eta}_{j}=+1$.

These conserved stabilizer sectors directly determine which bulk string correlators are fixed exactly in the ASPT state $|\rho^D\rangle\rangle$.

Because $C_{2m}^\eta=+1$, the following R\'{e}nyi-2 string correlator is exactly unity:
\begin{eqnarray}
&&\frac{\langle\langle\rho^D|S_{\mathrm{even}}(r)|\rho^D\rangle\rangle}{{\langle\langle\rho^D|\rho^D\rangle\rangle}}\equiv
\frac{\langle\langle\rho^D|\prod_{m=1}^{r}
C_{2m}^\eta|\rho^D\rangle\rangle}{{\langle\langle\rho^D|\rho^D\rangle\rangle}}\nonumber\\
&&=\frac{\langle\langle\rho^D|\eta^z_{1}\biggl(\prod_{m=1}^{r}\eta^x_{2m}\biggr)\eta^{z}_{2r+1}|\rho^D\rangle\rangle}{{\langle\langle\rho^D|\rho^D\rangle\rangle}}=1,
\label{R2string_eta_2m}
\end{eqnarray}
where 
$$S_{\mathrm{even}}(r)\equiv
Z_{1,u}Z_{1,\ell}
\left(\prod_{m=1}^{r} X_{2m,u}X_{2m,\ell}
\right)Z_{2r+1,u}Z_{2r+1,\ell}.
$$ 
In the $\eta$-spin representation, this is a conventional string correlator; in the original variables, it becomes
\begin{eqnarray}
\frac{\Tr[\rho^D S^{1/2}_{\mathrm{even}}(r)\rho^D S^{1/2}_{\mathrm{even}}(r)]}{\Tr[(\rho^D)^2]}=+1,
\end{eqnarray}
where $S^{1/2}_{\mathrm{even}}(r)\equiv Z_{1}
\left(\prod_{m=1}^{r} X_{2m}
\right)Z_{2r+1}$. 
This correlator was previously identified in Ref.~\cite{Lee2025}.

Likewise, $C^{\tau}_{2m+1}=+1$ gives a second R\'{e}nyi-2 string correlator:
\begin{eqnarray}
\frac{\langle\langle\rho^D|S_{\mathrm{odd}}(r)|\rho^D\rangle\rangle}{{\langle\langle\rho^D|\rho^D\rangle\rangle}}\equiv
\frac{\langle\langle\rho^D|\prod_{m=0}^{r-1}
C_{2m+1}^\tau|\rho^D\rangle\rangle}{{\langle\langle\rho^D|\rho^D\rangle\rangle}}=1,
\end{eqnarray}
where
$$
S_{\mathrm{odd}}(r)\equiv
Z_{0,u}Z_{0,l}
\left(\prod_{m=0}^{r-1} X_{2m+1,u}X_{2m+1,l}
\right)Z_{2r,u}Z_{2r,l}.
$$ 

The condition $C^{\eta}_{2m+1}=+1$ similarly yields the conventional string correlator on the lower copy,
\begin{eqnarray}
\frac{\langle\langle\rho^D|S^{1/2}_{\mathrm{odd},\ell}(r)|\rho^D\rangle\rangle}{\langle\langle\rho^D|\rho^D\rangle\rangle}=
\frac{\langle\langle\rho^D|\prod_{m=0}^{r-1}
C_{2m+1}^\eta|\rho^D\rangle\rangle}{{\langle\langle\rho^D|\rho^D\rangle\rangle}}=1,\nonumber\\
\label{S_odd_1_2}
\end{eqnarray}
where 
\begin{eqnarray}
S^{1/2}_{\mathrm{odd},\ell}(r)\equiv
Z_{0,\ell}
\left(\prod_{m=0}^{r-1} X_{2m+1,\ell}
\right)Z_{2r,\ell}.
\end{eqnarray}
In the original variables, this corresponds to the purity-weighted string correlator
\begin{eqnarray}
\frac{\Tr[(\rho^D)^2 S^{1/2}_{\mathrm{odd}}(r)]}{\Tr[(\rho^D)^2]}=+1.
\end{eqnarray}
Similar identifications were discussed in Refs.~\cite{Ma_PRXQuantum6,ma2024,Guo_2025,Guo2024_2}.

By contrast, because $C^{\tau}_{2m}$ is not conserved, the corresponding upper-copy string correlator is not fixed to unity:
\begin{eqnarray}
\frac{\langle\langle\rho^D|S^{1/2}_{\mathrm{even},u}(r)|\rho^D\rangle\rangle}{{\langle\langle\rho^D|\rho^D\rangle\rangle}}=
\frac{\langle\langle\rho^D|\prod_{m=1}^{r}
C_{2m}^\tau|\rho^D\rangle\rangle}{{\langle\langle\rho^D|\rho^D\rangle\rangle}}\neq 1,
\end{eqnarray}
where 
\begin{eqnarray}
S^{1/2}_{\mathrm{even},u}(r)\equiv
Z_{1,u}
\left(\prod_{m=1}^{r} X_{2m,u}
\right)Z_{2r+1,u}.
\end{eqnarray}
This follows from $[C^{\tau}_{2m},H_{\rm eff}]\neq 0$. Equivalently, in the original variables,
$$
\frac{\Tr[(\rho^D)^2 S^{1/2}_{\mathrm{even}}(r)]}{\Tr[(\rho^D)^2]}\neq 1.
$$
Thus, the ASPT does not possess an exact purity-weighted even-sublattice string correlator of this form.

Consequently, the $\tau$-$\eta$ spin mapping identifies which bulk string correlators are fixed by conserved stabilizers and which are not, providing a transparent bulk characterization of the cluster-state ASPT.

\section{Fate of the ASPT under the even-site $X$-decoherence channel}
We next examine the robustness of the ASPT under further symmetry-preserving decoherence. In the $\tau$-$\eta$ spin mapping, the ASPT is characterized by the conserved sectors $C^{\tau}_{2m+1}=+1$ and $C^{\eta}_{j}=+1$. We therefore ask how robust these constraints and their associated string orders remain when an additional quantum channel preserves the protecting strong and weak symmetries.

As a representative example, we consider the even-site $X$-decoherence channel. Its doubled-space representation is
\begin{eqnarray}
\hat{\mathcal{E}}_{X,e}
&=&
\prod^{L/2-1}_{m=0}\left[(1-p_x)+p_xX^*_{2m,u}X_{2m,\ell}\right]
\nonumber\\
&=&
(1-2p_x)^{L/4}
\exp\left[
\lambda_e\sum_m\eta^x_{2m}
\right],
\end{eqnarray}
with $\lambda_e=-\frac12\ln(1-2p_x)$.
This channel preserves both protecting symmetries, $S$ and $W$. In the doubled Hilbert space, its action is represented as a nonunitary filtering operation that produces the filtered state
\begin{eqnarray}
|\rho^D_X(\beta,p_x)\rangle\rangle=\hat{\mathcal{E}}_{X,e}|\rho^{D}(\beta)\rangle\rangle.
\end{eqnarray}

\subsection{Qualitative effective Hamiltonian}
To identify the competing orders, we first consider the following qualitative effective Hamiltonian based on $H_{\rm eff}$:
\begin{align}
H_{\rm ee}
=&-\frac{\cosh 2\beta}{2}
\sum_m C^\tau_{2m}\left(1+C^\eta_{2m}\right)\nonumber\\
&-\frac{1}{2}\sum_m C^\eta_{2m+1}\left(1+C^\tau_{2m+1}\right)
\nonumber\\
&-\frac{\sinh 4\beta}{2}
\sum_m \tau^z_{2m}
-h\sum_m \eta^x_{2m},
\label{H_ee}
\end{align}
where the last term represents the doubled-space filtering operation and $h$ is an effective parameter related to $\lambda_e$. The Hamiltonian $H_{\rm ee}$ preserves the strong and weak symmetries in Eqs.~(\ref{strong_sym1})--(\ref{weak_sym}). Although $H_{\rm ee}$ is only qualitative, its simple structure identifies the competing orders. In the sectors $C^{\tau}_{2m+1}=C^{\eta}_{2m}=+1$, it decomposes as $H_{\rm ee}=H_{\tau}+H_{\eta}$, where
\begin{align}
H_{\tau}
&=-\cosh 2\beta\sum_m C^\tau_{2m}-\frac{\sinh 4\beta}{2}\sum_m \tau^z_{2m},
\\
H_{\eta}
&=-\sum_m C^\eta_{2m+1}
-h\sum_m\eta^z_{2m-1}\eta^z_{2m+1}.
\end{align}
The $\eta$ sector has the form of a transverse-field Ising model (TFIM) \cite{PFEUTY197079,sachdev2011}: the filtering-induced Ising coupling competes with $C^\eta_{2m+1}$ and can suppress the string correlator in Eq.~(\ref{S_odd_1_2}). The corresponding long-range ordered regime would be characterized by
\begin{eqnarray}
&&\lim_{r\to\infty}
\frac{\langle\langle \rho^D_X(\beta,p_x)|
\eta^z_{2m+1}\eta^z_{2(m+r)+1}
|\rho^D_X(\beta,p_x)\rangle\rangle}
{\langle\langle \rho^D_X(\beta,p_x)|\rho^D_X(\beta,p_x)\rangle\rangle}
\nonumber\\
&&=
\lim_{r\to\infty}
\frac{
\Tr[
Z_{2m+1}Z_{2(m+r)+1}\rho^D_X
Z_{2m+1}Z_{2(m+r)+1}\rho^D_X
]}
{\Tr[(\rho^D_X)^2]}\nonumber\\
&&=\mathcal{O}(1).
\end{eqnarray}
A finite value of this R\'{e}nyi-2 $ZZ$ long-range correlator signals SWSSB because the odd sublattice carries the strong $Z_2$ symmetry generated by $S$. This mechanism is reminiscent of those discussed in Refs.~\cite{Tantivasadakarn2022,Lu2023_feedback,KOI_2024_gc,lu2026nonequilibrium,KO2026}. Thus, the qualitative effective Hamiltonian suggests that increasing the channel strength suppresses the ASPT string correlation in Eq.~(\ref{S_odd_1_2}) and enhances SWSSB correlations.
However, we do not infer from this Hamiltonian alone that the actual filtered state undergoes a TFIM-like phase transition from the ASPT to an SWSSB state. 
Rather, $H_{ee}$ should be regarded as a qualitative description of the competing orders. 
We therefore next examine the exact parent Hamiltonian associated with the filtered state.

\subsection{Exact parent Hamiltonian from $H_{\rm eff}$}
Beyond the qualitative effective Hamiltonian $H_{\rm ee}$, we apply the transformation scheme of Ref.~\cite{Lee2025} (see Appendix C) to construct an exact parent Hamiltonian whose ground state is $|\rho^D_X(\beta,p_x)\rangle\rangle$. In particular, we apply the transformation to the sector $H^{X_e}_{\rm eff}\equiv -\frac{1}{2}\sum_m C^\eta_{2m+1}(1+C^\tau_{2m+1})$ of $H_{\rm eff}$, whose ground state satisfies $C^{\eta}_{2m+1}=+1$. The transformed sector is
\begin{eqnarray}
H^{X_e,D}_{\rm eff}
&=&
\sum_{m}A^{F\dagger}_{2m+1}A^F_{2m+1}
+({\rm const.})
\label{HXeD_eff}
\end{eqnarray}
where $A^F_{2m+1}=\frac{1}{2}[1-e^{2\lambda_e(\eta^x_{2m}+\eta^x_{2m+2})}C^{\eta}_{2m+1}]$. 
Expanding this expression for small $\lambda_e$ gives
\begin{eqnarray}
H^{X_e,D}_{\rm eff}&=&
\sum^{L/2-1}_{m=0}
\biggl[
\frac{1-C^{\eta}_{2m+1}}{2}
-2\lambda_e\eta^{x}_{2m}\biggr]+O(\lambda_e^{2}).
\label{HXeD_eff_2}
\end{eqnarray}
The leading correction reproduces the $\sum_m\eta^x_{2m}$ term included in $H_{\rm ee}$, whereas the $O(\lambda_e^2)$ terms generate additional interactions that may be important for the robustness of the ASPT. Thus, $H_{\rm ee}$ identifies the competing orders but does not by itself establish a phase transition in the filtered state $\hat{\mathcal{E}}_{X,e}|\rho^D(\beta)\rangle\rangle$. The exact form of Eq.~(\ref{HXeD_eff_2}) is shown in the last part of Appendix C.

\subsection{Limit: $p_x\longrightarrow 1/2$}

As a final part of this section, we consider the maximal-decoherence limit
$p_x\longrightarrow 1/2$ of the channel $\hat{\mathcal{E}}_{X,e}$. In this limit,
\begin{eqnarray}
\lim _{p_x\to 1/2}\hat{\mathcal{E}}_{X,e}
&=&
\prod^{L/2-1}_{m=0}\biggl(\frac{1+\eta^x_{2m}}{2}\biggr).
\end{eqnarray}
Thus, the doubled-space representation of the channel becomes a projector onto the sector
$\eta^x_{2m}=+1$ for all even sites in the doubled system.
This limit therefore provides a simple endpoint picture complementary to the
qualitative effective-Hamiltonian and exact parent-Hamiltonian analyses discussed above.

In the projected sector, the surviving stabilizer constraints enforce the
long-range $ZZ$ correlation, and the decohered state exhibits the exact order
\begin{eqnarray}
&&\lim_{p_x\to 1/2}\lim_{r\to \infty}\langle \langle \rho^D_X(\beta,p_x)| Z_{2m+1,u}Z_{2m+1,\ell}\nonumber\\
&&\:\:\:\times Z_{2(m+r)+1,u}Z_{2(m+r)+1,\ell}
|\rho^D_X(\beta,p_x)\rangle\rangle\nonumber\\
&&\:\:\:/\langle\langle \rho^D_X(\beta,p_x)|\rho^D_X(\beta,p_x)\rangle\rangle
=+1,
\end{eqnarray}
corresponding to the exact SWSSB order. 
Thus, the maximal-decoherence limit
provides an analytically controlled endpoint of the deformation from the ASPT.

\section{Fate of the ASPT under symmetry-preserving no-click postselection}
We next consider a distinct symmetry-preserving operation that competes with the conserved quantity $C^\eta_{2m}$ and therefore probes the R\'{e}nyi-2 string correlator in Eq.~(\ref{R2string_eta_2m}). Specifically, we introduce a local weak $X$ measurement \cite{Biella2021} with Kraus operators
\begin{align}
M_{+,j}^{(X)}
&=
\sqrt{\frac{1+q_x}{2}}\,P_{+,j}^{(X)}
+
\sqrt{\frac{1-q_x}{2}}\,P_{-,j}^{(X)},\\
M_{-,j}^{(X)}
&=
\sqrt{\frac{1-q_x}{2}}\,P_{+,j}^{(X)}
+
\sqrt{\frac{1+q_x}{2}}\,P_{-,j}^{(X)},
\end{align}
where $M_{+,j}^\dagger M_{+,j}=
\frac{1+q_xX_j}{2}$, $M_{-,j}^\dagger M_{-,j}=
\frac{1-q_xX_j}{2}$,  $P_{\pm,j}^{(X)}=\frac{1\pm X_j}{2}$ and $M_{+,j}^\dagger M_{+,j}
+M_{-,j}^\dagger M_{-,j}
=I$. The measurement strength satisfies $0\le q_x<1$, with the projective limit reached as $q_x\to1$.

We postselect the $+$ outcome at every odd site, thereby defining the no-click postselection \cite{Gopalakrishnan2021,Biella2021},
\begin{eqnarray}
\rho^{{\rm ps}}(q_x)
=\frac{
K_{+,o}^{(q_x)}\rho^D K_{+,o}^{(q_x)\dagger}
}{{\rm Tr}[K_{+,o}^{(q_x)}\rho^D K_{+,o}^{(q_x)\dagger}]}, 
\end{eqnarray}
where $K_{+,o}^{(q_x)}=\prod^{L/2-1}_{m=0}M_{+,2m+1}^{(X)}$. 
This postselection preserves both protecting symmetries, $S$ and $W$. The local Kraus operator can be written in the filtering form
\begin{align}
M_{+,2m+1}^{(X)}=C_{q_x} e^{\kappa_{q_x} X_{2m+1}},
\end{align}
where $C_{q_x}=\frac{(1-q_x^2)^{1/4}}{\sqrt{2}}$ and
$\kappa_{q_x}=\frac{1}{4}\ln\frac{1+q_x}{1-q_x}$. Up to normalization, the corresponding doubled state is
\begin{eqnarray}
&&|\rho^{{\rm ps}}(q_x)\rangle\rangle
\propto K_{+,o,u}^{(q_x)*}K_{+,o,\ell}^{(q_x)}|\rho^D(\beta)\rangle\rangle\nonumber\\
&&=\exp\left[\kappa_{q_x}\sum^{L/2-1}_{m=0} \eta^x_{2m+1}(1+\tau^x_{2m+1})\right]
|\rho^D(\beta)\rangle\rangle,
\end{eqnarray}
where we used the $\tau$-$\eta$ spin mapping. We leave this state unnormalized because all observables considered below are divided by $\langle\langle\rho^{{\rm ps}}(q_x)|\rho^{{\rm ps}}(q_x)\rangle\rangle$.

\subsection{Qualitative effective Hamiltonian}
As in the decoherence case, we first use the following qualitative effective Hamiltonian based on $H_{\rm eff}$:
\begin{eqnarray}
H_{\rm ee}^{\rm ps}&=&-\frac{\cosh 2\beta}{2}\sum_m C^\tau_{2m}(1+C^\eta_{2m})\nonumber\\
&&-\frac{1}{2}\sum_m C^\eta_{2m+1}(1+C^\tau_{2m+1})
-\frac{\sinh 4\beta}{2}\sum_m\tau^z_{2m}\nonumber\\
&&-h_{\rm ps}\sum_m\eta^x_{2m+1}(1+\tau^x_{2m+1})
+({\rm const.}).
\label{H_ps}
\end{eqnarray}
Here, $h_{\rm ps}$ is an effective parameter related to $\kappa_{q_x}$. We decouple this Hamiltonian using a mean-field approximation, as detailed in Appendix D. The resulting mean-field Hamiltonian is
$H^{\rm ps}_{\rm MF}\equiv H^{\rm MF}_{\tau}+H^{\rm MF}_{\eta}$. 
The $\tau$ sector is given by
\begin{eqnarray}
H_{\tau}^{\rm MF}
&=&-\frac{\cosh 2\beta}{2}
(1+\langle C^\eta_{2m}\rangle)\sum_m C^\tau_{2m}
-\frac{\sinh 4\beta}{2}\sum_m\tau^z_{2m}\nonumber\\
&&-h_{\rm ps}\langle\eta^z_{2m}\eta^z_{2m+2}\rangle
\sum_m\tau^z_{2m}\tau^z_{2m+2}
+({\rm const.}),
\label{MFH_1}
\end{eqnarray}
while the $\eta$ sector is
\begin{eqnarray}
H_{\eta}^{\rm MF}
&=&-\frac{\cosh 2\beta}{2}\langle C^\tau_{2m}\rangle
\sum_m C^\eta_{2m}\nonumber\\
&&-h_{\rm ps}
(1+\langle\tau^z_{2m}\tau^z_{2m+2}\rangle)
\sum_m\eta^z_{2m}\eta^z_{2m+2}
+({\rm const.}).\nonumber\\
\label{MFH_2}
\end{eqnarray}
Here, we fix the sector $C^\tau_{2m+1}=C^\eta_{2m+1}=+1$, and $\langle C^{\tau(\eta)}_{2m}\rangle$, $\langle\tau^z_{2m}\tau^z_{2m+2}\rangle$, and $\langle\eta^z_{2m}\eta^z_{2m+2}\rangle$ are mean fields. The $\eta$ sector exhibits a competition between the $C^\eta_{2m}$ term and the postselection-induced coupling $\eta^z_{2m}\eta^z_{2m+2}$. The mean-field description therefore suggests a possible Ising-like instability at finite postselection strength. Correspondingly, increasing $h_{\rm ps}$ suppresses the R\'{e}nyi-2 string correlator in Eq.~(\ref{R2string_eta_2m}) and enhances the long-range correlation
\begin{eqnarray}
&&\lim_{r\to\infty}
\frac{\langle\langle \rho^{{\rm ps}}(q_x) |\eta^z_{2m}\eta^z_{2(m+r)}|\rho^{{\rm ps}}(q_x)\rangle\rangle}
{\langle\langle \rho^{{\rm ps}}(q_x)|\rho^{{\rm ps}}(q_x)\rangle\rangle}\nonumber\\
&&=
\lim_{r\to\infty}
\frac{\langle\langle \rho^{{\rm ps}}(q_x)|Z_{2m,\ell}Z_{2(m+r),\ell}|\rho^{{\rm ps}}(q_x)\rangle\rangle}
{\langle\langle \rho^{{\rm ps}}(q_x)|\rho^{{\rm ps}}(q_x)\rangle\rangle}
\nonumber\\
&&=
\lim_{r\to\infty}
\frac{{\rm Tr}\left[(\rho^{\rm ps}(q_x))^2 Z_{2m}Z_{2(m+r)}\right]}
{{\rm Tr}[(\rho^{\rm ps}(q_x))^2]}
=\mathcal{O}(1).
\end{eqnarray}
This is a purity-weighted $ZZ$ correlator. Because $Z_j$ is charged under the weak symmetry generated by $W$, a nonzero asymptotic value characterizes WSSB on the even sublattice \cite{Ando2026}. Thus, the qualitative effective Hamiltonian identifies the competition between ASPT and WSSB orders, but it does not by itself establish a phase transition in the actual postselected state.

\subsection{Exact parent-Hamiltonian transformation}
Beyond the qualitative effective Hamiltonian $H^{\rm ps}_{\rm ee}$, we apply the transformation scheme of Ref.~\cite{Lee2025} (see Appendix C) to construct the exact parent Hamiltonian whose ground state is $|\rho^{{\rm ps}}(q_x)\rangle\rangle$.

We begin with the pure cluster Hamiltonian in the doubled space,
\begin{eqnarray}
H^d_{c}=\sum_{j,\alpha=u,\ell}A^{(0)\dagger}_{j,\alpha}A^{(0)}_{j,\alpha} +({\rm const.}),
\end{eqnarray}
where $A^{(0)}_{j,\alpha}\equiv \frac{1-K_{j,\alpha}}{2}$. 
Following Ref.~\cite{Lee2025}, we denote the doubled-space no-click filter by $\hat{M}^{(X)}_+$.
The transformed annihilation operator under $\hat{\mathcal E}_{Z,e}$ and $\hat{M}^{(X)}_+$ is
\begin{eqnarray}
&&A^{D,{\rm ps}}_{2m,u}
=
\hat{M}^{(X)}_+\hat{\mathcal E}_{Z,e}
A^{(0)}_{2m,u}
\hat{\mathcal E}_{Z,e}^{-1}
(\hat{M}^{(X)}_+)^{-1}
\nonumber\\
&&=
\frac{1}{2}\left[
1-e^{2\beta Z_{2m,u}Z_{2m,\ell}}
e^{2\kappa_{q_x}(X_{2m-1,u}+X_{2m+1,u})}
K_{2m,u}
\right].\nonumber\\
\end{eqnarray}
The resulting exact parent Hamiltonian is 
\begin{eqnarray}
H^{\rm ps}_{\rm eff}
&=&\sum_{m,\alpha=u,\ell}\biggl[A^{(0)\dagger}_{2m+1,\alpha}A^{(0)}_{2m+1,\alpha} +A^{D,{\rm ps}\dagger}_{2m,\alpha}A^{D,{\rm ps}}_{2m,\alpha}\biggr]\nonumber\\
&&+({\rm const.})\nonumber\\
&=&
H_{\rm eff}+\delta H^{(1)}_{\rm ps}+O(\kappa_{q_x}^2)+({\rm const.}),
\end{eqnarray} 
where 
\begin{eqnarray}
\delta H^{(1)}_{\rm ps}
&=&
-2\kappa_{q_x}\cosh 4\beta
\sum_m
\eta^x_{2m+1}\left(1+\tau^x_{2m+1}\right)
\nonumber\\
&&
+\kappa_{q_x}\sinh 4\beta
\sum_m
\tau^z_{2m}[
\eta^x_{2m-1}(1+\tau^x_{2m-1})\nonumber\\
&&+\eta^x_{2m+1}(1+\tau^x_{2m+1})]
\nonumber\\
&&
-\kappa_{q_x}\sinh 2\beta
\sum_m
\tau^z_{2m}C^\tau_{2m}
[\eta^x_{2m-1}(\tau^x_{2m-1}+C^\eta_{2m})\nonumber\\
&&+\eta^x_{2m+1}(\tau^x_{2m+1}+C^\eta_{2m})
].
\end{eqnarray}
The $O(\kappa^2_{q_x})$ terms generate further interactions that may significantly affect the fate of the ASPT. Consequently, although the qualitative effective Hamiltonian identifies the competing orders, it does not establish a phase transition in the no-click postselected state $K^{(q_x)}_{+,o}|\rho^D(\beta)\rangle\rangle$.
In particular, the additional interactions appearing at the same leading order prevent us from identifying the Ising-like instability of the simplified description with an actual phase transition of the postselected state.
\begin{figure*}[t]
\begin{center}
\includegraphics[width=17cm]{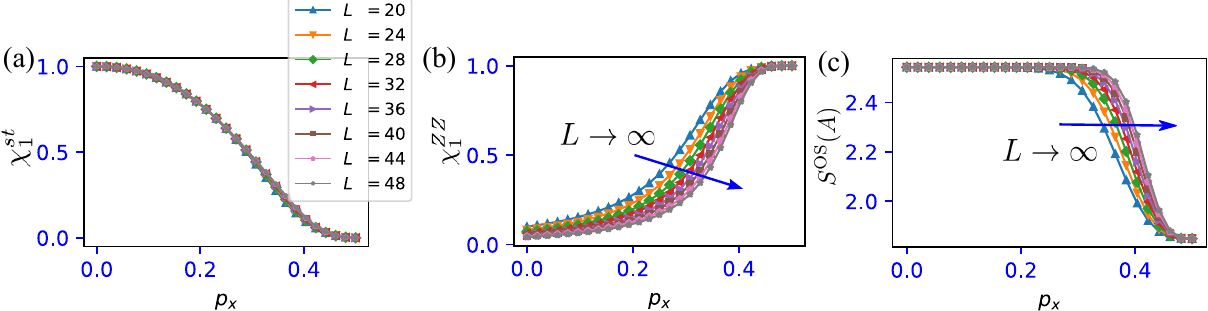}
\end{center}
\vspace{-0.5cm}
\caption{$p_x$ dependence of (a) the normalized sum of the purity-weighted string correlator $\chi^{st}_1$, (b) the normalized sum of the R\'{e}nyi-2 $ZZ$ correlator $\chi^{ZZ}_1$, and (c) the OSEE $S^{\rm OS}(A)$. The arrows in (b) and (c) indicate the evolution of the curves with increasing system size $L$.}
\label{Fig1}
\end{figure*}

\begin{figure}[t]
\begin{center}
\includegraphics[width=8.5cm]{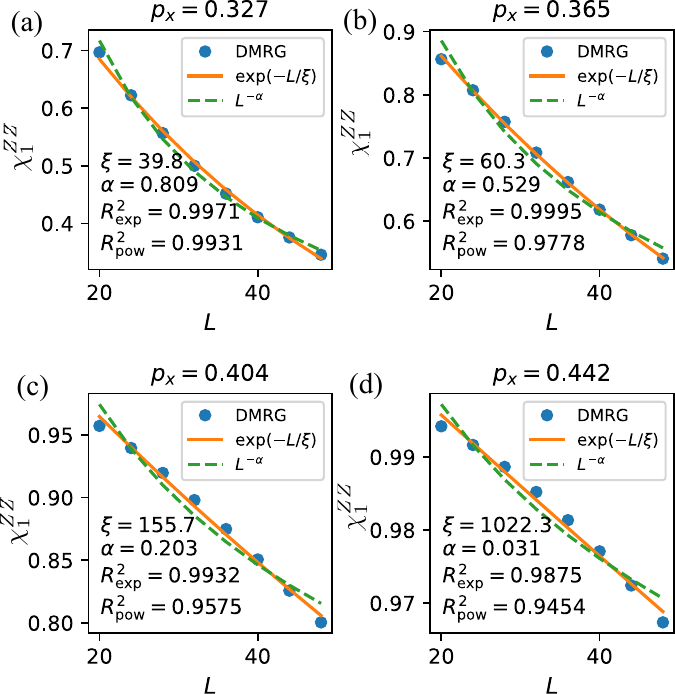}
\end{center}
\vspace{-0.5cm}
\caption{System-size dependence of the normalized R\'{e}nyi-2 $ZZ$ correlator $\chi^{ZZ}_1$ for representative values of $p_x$: (a) $p_x=0.327$, (b) $p_x=0.365$, (c) $p_x=0.404$, and (d) $p_x=0.442$. 
The solid orange curves show the best exponential fits, and the dashed green curves show the best power-law fits. Here, $R^2_{\rm exp}$ and $R^2_{\rm pow}$ denote the coefficients of determination for the exponential and power-law fits, respectively.
Over the accessible system sizes, the exponential form provides the better fit.}
\label{Fig2}
\end{figure}
\subsection{Limit: $q_x\longrightarrow 1$}

As a final part of this section, we consider the projective limit
$q_x\longrightarrow 1$ of the no-click postselection $K^{(q_x)}_{+,o}$. In this limit,
\begin{eqnarray}
&&\lim_{q_x\to 1}K_{+,o,u}^{(q_x)*}K_{+,o,\ell}^{(q_x)}
=
\prod^{L/2-1}_{m=0}P_{+,2m+1,u}^{(X)*}P_{+,2m+1,\ell}^{(X)}
\nonumber\\
&&=
\prod^{L/2-1}_{m=0}\frac{1}{4}
(1+\tau^x_{2m+1}\eta^x_{2m+1})
(1+\eta^x_{2m+1}).
\end{eqnarray}
Thus, the no-click postselection becomes a projector onto the sector
$\tau^x_{2m+1}=\eta^x_{2m+1}=+1$ for all odd sites in the doubled system.
This limit therefore provides a simple endpoint picture complementary to the
qualitative effective-Hamiltonian and exact parent-Hamiltonian analyses discussed above.

In the projected sector, the corresponding long-range order becomes exact,
and the postselected state satisfies
\begin{eqnarray}
\lim_{q_x\to1}\lim_{r\to\infty}
\frac{\langle\langle \rho^{{\rm ps}}(q_x) |\eta^z_{2m}\eta^z_{2(m+r)}|\rho^{{\rm ps}}(q_x)\rangle\rangle}
{\langle\langle \rho^{{\rm ps}}(q_x)|\rho^{{\rm ps}}(q_x)\rangle\rangle}
=+1,
\end{eqnarray}
corresponding to the exact WSSB order. Thus, the projective limit provides a simple analytically controlled endpoint of the postselected ASPT state.

\section{Numerical investigation}
We now numerically investigate the robustness of the ASPT. We first prepare the doubled cluster-SPT ground state
$|\rho_{0}\rangle\rangle\equiv |\psi_0^*\rangle_u\otimes|\psi_0\rangle_\ell$,
where $|\psi_0\rangle$ is the ground state of the cluster chain governed by $H_{\rm c}$. We obtain $|\psi_0\rangle$ by DMRG using the TeNPy package \cite{
TeNPy_Lec,10.21468/SciPostPhysCodeb.41}. We use a maximum bond dimension $D=200$--$260$, discard singular values below $O(10^{-8})$, and impose an energy-convergence criterion $\Delta E<O(10^{-6})$. We then construct the ASPT state as $|\rho^D(\beta)\rangle\rangle\equiv\hat{\mathcal{E}}_{Z,e}|\rho_0\rangle\rangle$ using MPS filtering \cite{Haegeman2015,Orito2025,KOI2025_v2}. 
Finally, we apply either $\hat{\mathcal{E}}_{X,e}$ or $K_{+,o,u}^{(q_x)*}K_{+,o,\ell}^{(q_x)}$ to $|\rho^D(\beta)\rangle\rangle$ and evaluate the observables defined below.

\subsection{Even-site $X$-decoherence channel}
We first investigate the filtered ASPT state $|\rho^D_X\rangle\rangle\equiv\hat{\mathcal{E}}_{X,e}(p_x)|\rho^D(\beta)\rangle\rangle$.

We calculate three quantities. The first is the normalized sum of the purity-weighted string correlator,
\begin{eqnarray}
\chi^{st}_1=\frac{2}{L}\sum^{L/2}_{r=1}\frac{\langle\langle\rho^D_X|S^{1/2}_{\mathrm{even},u}(r)|\rho^D_X\rangle\rangle}{\langle \langle \rho^D_X
|\rho^D_X\rangle\rangle},
\end{eqnarray}
which probes the ASPT string order. The second is the normalized sum of the R\'{e}nyi-2 $ZZ$ correlator $\eta^z_{2m+1}\eta^z_{2(m+r)+1}$,
\begin{eqnarray}
\chi^{ZZ}_1=\frac{2}{L}\sum^{L/2}_{r=1}\frac{\langle \langle \rho^D_X| \eta^z_{2m+1}\eta^z_{2(m+r)+1}
|\rho^D_X\rangle\rangle}{\langle \langle \rho^D_X
|\rho^D_X\rangle\rangle},
\end{eqnarray}
which probes SWSSB correlations on the odd sublattice. 
The third is the operator-space entanglement entropy (OSEE) \cite{Prosen2007,Pizorn2009,Znidaric2008,Nieuwenburg2014,Nieuwenburg2018}. In the Choi representation, the OSEE is the bipartite entanglement entropy of the normalized doubled state (see Appendix E),
\begin{eqnarray}
S^{\rm OS}(A)=-{\rm Tr}[{\tilde \rho}_A \log {{\tilde \rho}_A}],
\end{eqnarray} 
where ${\tilde \rho}_A$ is obtained from ${\tilde \rho}\equiv |{\tilde \rho}^D\rangle\rangle\langle\langle {\tilde \rho}^D|$, with $|{\tilde \rho}^D\rangle\rangle=|\rho^D\rangle\rangle/\sqrt{\langle\langle \rho^D|\rho^D\rangle\rangle}$, by tracing out the complement of $A$. The OSEE diagnoses changes in the doubled-state correlation structure \cite{Prosen_PhysRevLett,Pizorn2009,Wellnitz2022,Orito2025,Ando2026,YK2026}. We choose $A$ as the left spatial half of the ladder, including both upper and lower degrees of freedom.

We set $p_z=0.2$, corresponding to $\beta\simeq0.2554$, and show the $p_x$ dependence of the three quantities in Fig.~\ref{Fig1}.
As shown in Fig.~\ref{Fig1}(a), the normalized purity-weighted string correlator $\chi^{\rm st}_{1}$ remains close to unity in the weak-decoherence regime and gradually decreases with increasing $p_x$, approaching zero toward the maximal-decoherence limit $p_x=1/2$. In contrast, the normalized R\'enyi-2 $ZZ$ correlator $\chi^{ZZ}_{1}$ in Fig.~\ref{Fig1}(b) increases with $p_x$ and approaches unity near $p_x=1/2$. This behavior is consistent with the analytical result in Sec.~IV\,C, where the doubled-space representation of the channel becomes a projector onto the $\eta^x_{2m}=+1$ sector and enforces the $ZZ$ correlation associated with SWSSB. Thus, the two quantities demonstrate the competition between ASPT string order and SWSSB correlations. The OSEE in Fig.~\ref{Fig1}(c) remains almost unchanged throughout the weak- and intermediate-decoherence regimes and decreases substantially only in the strong $p_x$-decoherence region. As indicated by the arrows in Figs.~\ref{Fig1}(b) and \ref{Fig1}(c), increasing $L$ shifts the pronounced growth of $\chi^{ZZ}_1$ and the suppression of the OSEE toward stronger decoherence, consistent with a strong finite-size crossover near the maximal-decoherence limit. 

\begin{figure*}[t]
\begin{center}
\includegraphics[width=17cm]{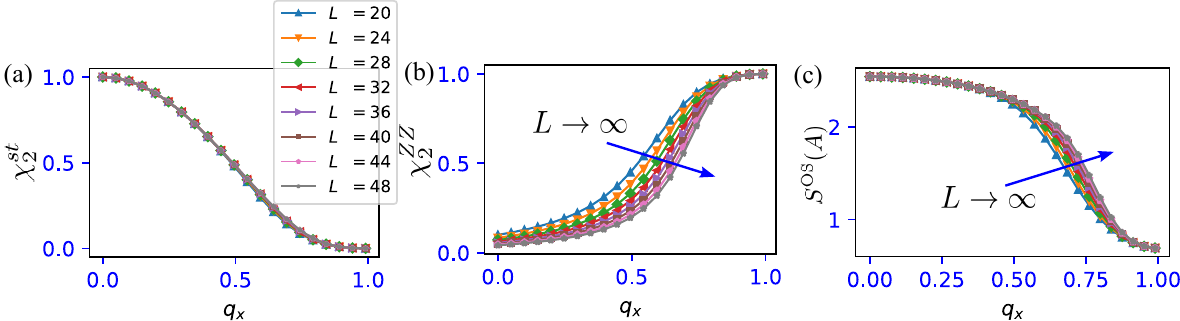}
\end{center}
\vspace{-0.5cm}
\caption{$q_x$ dependence of (a) the normalized sum of the R\'{e}nyi-2 string correlator $\chi^{st}_2$, (b) the normalized sum of the purity-weighted $ZZ$ correlator $\chi^{ZZ}_2$, and (c) the OSEE $S^{\rm OS}(A)$. The arrows in (b) and (c) indicate the evolution with increasing $L$, showing that the pronounced changes shift toward the nearly projective regime.}
\label{Fig3}
\end{figure*}
\begin{figure}[t]
\begin{center}
\includegraphics[width=8.5cm]{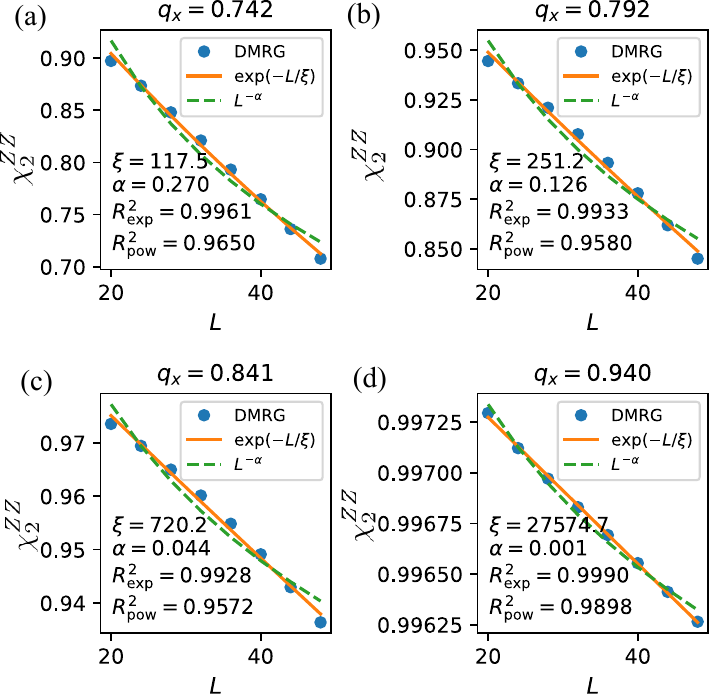}
\end{center}
\vspace{-0.5cm}
\caption{System-size dependence of the normalized purity-weighted $ZZ$ correlator $\chi^{ZZ}_2$ for representative values of $q_x$: (a) $q_x=0.742$, (b) $q_x=0.792$, (c) $q_x=0.841$, and (d) $q_x=0.940$. The solid orange curves show the best exponential fits, and the dashed green curves show the best power-law fits. 
Here, $R^2_{\rm exp}$ and $R^2_{\rm pow}$ denote the coefficients of determination for the exponential and power-law fits, respectively.
Over the accessible system sizes, the exponential form provides the better fit.}
\label{Fig4}
\end{figure}
Moreover, to determine whether the growth of $\chi^{ZZ}_{1}$ corresponds to a genuine finite-$p_x$ ordered phase, we examine its system-size dependence in Figs.~\ref{Fig2}(a)--\ref{Fig2}(d). For representative values of $p_x$, we compare exponential and power-law fits. The numerical data are better described by the exponential form over the available system sizes.
At the same time, the extracted characteristic length grows rapidly with increasing $p_x$, from $\xi\simeq 40$ at $p_x=0.327$ to $\xi\simeq 10^{3}$ at $p_x=0.442$. 
This rapid increase explains the apparently sharp development of the $ZZ$ correlation seen in Fig.~\ref{Fig1}(b). 
Nevertheless, the present finite-size behavior does not provide clear evidence for a finite-$p_x$ critical point. 
Rather, it is consistent with an increasingly large characteristic length as the maximal-decoherence limit $p_x=1/2$ is approached, although the available system sizes do not allow us to determine its asymptotic behavior. At this limit, the SWSSB order becomes exact.

This behavior also clarifies the relation to the qualitative effective Hamiltonian discussed in Sec.~IV\,A. The Hamiltonian $H_{\rm ee}$ contains an explicit TFIM sector and therefore suggests a finite-$p_x$ Ising transition between the ASPT and SWSSB regimes.
However, such a transition is not supported by the present numerical results for the actual filtered state. 
Instead, the numerical results indicate a strong finite-size crossover toward the SWSSB state as $p_x$ approaches the maximal-decoherence limit. 
This difference is consistent with the exact parent-Hamiltonian analysis in Sec.~IV\,B, where additional interaction terms beyond the simple TFIM structure are generated by the filtering transformation.
Thus, the TFIM Hamiltonian should be regarded as a qualitative description identifying the competing ASPT and SWSSB orders, rather than as a direct prediction of the critical behavior of the filtered state. These results indicate that the ASPT remains robust over a broad range of finite decoherence strengths, while the pronounced reconstruction toward SWSSB occurs only close to the maximal-decoherence limit.

\subsection{No-click postselection}
Next, we numerically investigate the no-click postselected state given by $|\rho^{{\rm ps}}(q_x)\rangle\rangle \propto K_{+,o,u}^{(q_x)*}K_{+,o,\ell}^{(q_x)}|\rho^D(\beta)\rangle\rangle$. 

We again calculate three quantities. The first is the normalized sum of the R\'enyi-2 string correlator,
\begin{eqnarray}
\chi^{st}_2=\frac{2}{L}\sum^{L/2}_{r=1}\frac{\langle\langle\rho^{{\rm ps}}(q_x)|S_{\mathrm{even}}(r)|\rho^{{\rm ps}}(q_x)\rangle\rangle}{\langle \langle \rho^{{\rm ps}}(q_x)
|\rho^{{\rm ps}}(q_x)\rangle\rangle},
\end{eqnarray}
which probes the ASPT string order. The second is the normalized sum of the purity-weighted $ZZ$ correlator $\eta^z_{2m}\eta^z_{2(m+r)}$,
\begin{eqnarray}
\chi^{ZZ}_2=\frac{2}{L}\sum^{L/2}_{r=1}\frac{\langle \langle \rho^{{\rm ps}}(q_x)| \eta^z_{2m}\eta^z_{2(m+r)}
|\rho^{{\rm ps}}(q_x)\rangle\rangle}{\langle \langle \rho^{{\rm ps}}(q_x)
|\rho^{{\rm ps}}(q_x)\rangle\rangle},
\end{eqnarray}
which probes WSSB correlations on the even sublattice. The third is the OSEE defined in the previous subsection.

We again set $p_z=0.2$, corresponding to $\beta\simeq 0.2554$. The $q_x$ dependence of the three quantities is shown in Fig.~\ref{Fig3}.
As shown in Fig.~\ref{Fig3}(a), the total sum of the R\'enyi-2 string correlation $\chi^{st}_2$ decreases continuously with increasing $q_x$ and becomes strongly suppressed only close to the projective limit $q_x=1$.
In contrast, the normalized purity-weighted $ZZ$ correlator $\chi^{ZZ}_2$ in Fig.~\ref{Fig3}(b) grows monotonically and approaches unity as $q_x\rightarrow 1$. This behavior is consistent with the analytical result in Sec.~V C, where the no-click postselection reduces to a projector onto the sector $\tau^x_{2m+1}=\eta^x_{2m+1}=+1$. The projective limit therefore realizes exact WSSB order on the even sublattice.
The OSEE shown in Fig.~\ref{Fig3}(c) also decreases with increasing $q_x$, but its suppression is gradual over a broad range of $q_x$ and becomes particularly strong only in the nearly projective regime.
These results indicate that the ASPT correlation structure remains robust against finite-strength no-click postselection. As indicated by the arrows in Figs.~\ref{Fig3}(b) and \ref{Fig3}(c), the characteristic changes in both $\chi^{ZZ}_2$ and the OSEE shift toward larger $q_x$ with increasing $L$, consistent with a strong finite-size crossover concentrated near the projective limit.

We further examine the system-size dependence of the $ZZ$ correlation at finite $q_x$ in Fig.~\ref{Fig4}.
For representative values of $q_x$, the numerical data are compared with exponential and power-law fittings.
Over the accessible system sizes, the exponential form gives a better description of the data.
At the same time, the extracted characteristic length rapidly increases as $q_x$ approaches unity, from $\xi\simeq 1.2\times10^{2}$ at $q_x=0.742$ to $\xi\simeq 2.8\times10^{3}$ at $q_x=0.940$.
Such a rapid growth accounts for the apparently sharp increase of the $ZZ$ correlation in Fig.~\ref{Fig3}(b).
Nevertheless, the present finite-size behavior does not provide clear evidence for a finite-$q_x$ critical point.  
Rather, it is consistent with a rapidly increasing characteristic length as the projective limit $q_x\to 1$ is approached, although the available system sizes do not allow us to determine its asymptotic behavior.

This behavior also clarifies the relation to the mean-field description discussed in Sec.~V A.
The mean-field Hamiltonian contains an Ising-like competition between the ASPT stabilizer term and the postselection-induced $\eta^z\eta^z$ coupling, and therefore suggests a possible finite-$q_x$ transition from the ASPT to the WSSB regime. However, no such transition is supported by the present numerical results for the actual postselected state.
Instead, the numerical data indicate a strong finite-size crossover toward the WSSB state as $q_x$ approaches the projective limit.
This discrepancy is consistent with the exact parent-Hamiltonian analysis in Sec.~V B, where the no-click filtering generates several additional correlated interactions beyond the simple mean-field Ising structure already at the leading order.
These additional terms substantially modify the simplified mean-field picture.
Thus, the mean-field Hamiltonian should be regarded as a qualitative description identifying the competing ASPT and WSSB orders, rather than as a direct prediction of the critical behavior.
The numerical results therefore support a picture in which the ASPT remains robust over a broad finite-postselection regime and evolves toward the WSSB state through a strong finite-size crossover, with exact WSSB order realized only in the no-click projective limit $q_x\to 1$.

\section{Conclusion}
In this study, we investigated the fate of the one-dimensional cluster-state ASPT under symmetry-preserving quantum operations using the doubled-Hilbert-space formalism. By introducing two effective spins, $\tau$ and $\eta$, on each doubled rung, we obtained a transparent representation of the ASPT parent Hamiltonian and identified its exact local conserved quantities. This representation establishes a direct correspondence between the conserved quantities and the different purity-weighted and R\'enyi-2 string orders used to characterize ASPT states. Thus, the $\tau$-$\eta$ spin mapping provides a unified microscopic picture of the order structure of the cluster-state ASPT.

We further studied the deformation of the ASPT under two symmetry-preserving operations. 
First, we considered the even-site $X$-decoherence channel. The qualitative effective Hamiltonian contains a TFIM sector, indicating competition between the ASPT string order and the R\'enyi-2 $ZZ$ long-range correlations associated with SWSSB. The exact parent-Hamiltonian construction confirms the corresponding leading perturbation while also generating additional correlated terms that can substantially modify the Ising-like instability suggested by the qualitative effective Hamiltonian. In the maximal-decoherence limit, the doubled-space representation of the channel becomes a projector, and exact SWSSB order follows directly from the projector constraints. Second, we considered symmetry-preserving no-click postselection. Although a mean-field treatment of the qualitative effective Hamiltonian suggests an Ising-like instability toward WSSB, an expansion of the exact parent Hamiltonian shows that several correlated interactions appear at the same leading order. Thus, the mean-field instability does not by itself establish a phase transition in the actual postselected state. 

Our DMRG and MPS filtering calculations show  that the ASPT remains robust over a broad range of finite decoherence and postselection strengths. SWSSB and WSSB correlations become pronounced only near the maximal-decoherence and projective limits, respectively, through strong finite-size crossovers. Exact symmetry-breaking order is realized at the corresponding limiting points. Thus, rather than being immediately destroyed by symmetry-preserving nonunitary operations, the ASPT is continuously deformed over a broad parameter regime before reaching the symmetry-breaking endpoints.

An interesting extension may be to higher-dimensional cluster states, where decoherence can lead to ASPT-like structures protected in part by higher-form symmetries \cite{Lee2025,Guo-and-Ashida2024,lee2025robust}. 
Understanding how higher-dimensional ASPT states change under symmetry-preserving decoherence channels is an interesting direction for future work. Extending the present microscopic spin-mapping approach to such systems is nontrivial because the relevant conserved structures involve extended loop operators rather than the local stabilizer sectors appearing in one dimension.

\section*{Acknowledgements}
This work is supported by JSPS KAKENHI: 26K06956(Y.K.) and 26K17056(T.O.). 

\section*{Data availability}
The data that support the findings of this article are not publicly available. The data are available from the authors upon reasonable request.


\appendix

\section{Choi mapping}
Let $\rho$ be a density matrix acting on $\mathcal{H}$. In the Choi mapping \cite{Choi1975,JAMIOLKOWSKI1972}, it is vectorized in the doubled Hilbert space $\mathcal{H}_{u}\otimes\mathcal{H}_{\ell}$ as
\begin{eqnarray}
|\rho\rangle\rangle
&\equiv&
\sum_{i,j}\rho_{ji}|i\rangle_u\otimes|j\rangle_\ell
=\sum_i|i\rangle_u\otimes\rho|i\rangle_\ell,
\end{eqnarray}
where $u$ and $\ell$ carry the bra and ket indices, respectively. The bases $\{|i\rangle_u\}$ and $\{|j\rangle_\ell\}$ are orthonormal bases of the two copies of $\mathcal{H}$. With this convention,
\begin{eqnarray}
\rho_0=|\psi_0\rangle\langle\psi_0|
\quad\longrightarrow\quad
|\rho_0\rangle\rangle
=|\psi_0^*\rangle_u\otimes|\psi_0\rangle_\ell,
\end{eqnarray}
and $\langle\langle\rho|\rho\rangle\rangle=\Tr[\rho^\dagger\rho]=\Tr[\rho^2]$ for a Hermitian density matrix.

A quantum channel in Kraus form is written as \cite{Nielsen2011,lidar2020}
\begin{eqnarray}
\hat{\mathcal{E}}[\rho]=\sum^{M-1}_{\alpha=0}\hat{E}_{\alpha}\rho \hat{E}^\dagger_{\alpha},
\label{Kraus_sum}
\end{eqnarray} 
where the Kraus operators satisfy $\sum^{M-1}_{\alpha=0}\hat{E}^\dagger_{\alpha}\hat{E}_{\alpha}=I$. Under the above vectorization convention, the channel is represented by the doubled-space operator \cite{Grover2024,Lee2025}
\begin{eqnarray}
\hat{\mathcal{E}}\longrightarrow \hat{\mathcal{E}}
=\sum^{M-1}_{\alpha=0}\hat{E}^*_{\alpha,u}\otimes \hat{E}_{\alpha,\ell},
\end{eqnarray}
which acts as $|\mathcal{E}(\rho)\rangle\rangle=\hat{\mathcal{E}}|\rho\rangle\rangle$. Although $\hat{\mathcal{E}}$ is generally nonunitary, the physical channel $\hat{\mathcal{E}}$ is completely positive and trace preserving. The action of $\hat{\mathcal{E}}$ can be efficiently implemented as an MPS filtering operation \cite{Haegeman2015,Orito2025,KOI2025_v2}.

Physical correlators can then be expressed as expectation values in the doubled Hilbert space \cite{Ma_PRXQuantum6}. For Hermitian observables $O_1$ and $O_2$,
\begin{eqnarray}
\frac{\Tr[O_1\rho O_2\rho]}{\Tr[\rho^2]}
&=&
\frac{\langle\langle\rho|O^*_{1,u}O_{2,\ell}|\rho\rangle\rangle}
{\langle\langle\rho|\rho\rangle\rangle},
\end{eqnarray}
whereas the purity-weighted correlator is \cite{Ando2026}
\begin{eqnarray}
\frac{\Tr[O_1\rho^2]}{\Tr[\rho^2]}
&=&
\frac{\langle\langle\rho|O^*_{1,u}|\rho\rangle\rangle}
{\langle\langle\rho|\rho\rangle\rangle}
=
\frac{\langle\langle\rho|O_{1,\ell}|\rho\rangle\rangle}
{\langle\langle\rho|\rho\rangle\rangle}.
\end{eqnarray}

\section{Strong and weak symmetries}
In this Appendix, we briefly summarize two types of notions of symmetries for mixed states, namely, strong and weak symmetries \cite{Buca2012,groot2022}.
This work picks up the on-site unitary $Z_2$ symmetry group, the generator of which is $\{\hat{1},U_{Z_2}\}$ with $U^2_{Z_2}=\hat{1}$. 

First, the strong symmetry of a density matrix is defined as 
$$
U_{Z_2}\rho=e^{i\theta}\rho,\;\;\; \rho U^{\dagger}_{Z_2}=e^{-i\theta}\rho,
$$ 
where $\rho$ is a mixed state and $\theta$ is a global phase factor. 

The second notion is a weak symmetry. This is defined as  
$$
U_{Z_2}\rho U^\dagger_{Z_2} = \rho.
$$ 
The symmetry is satisfied on ensemble averages \cite{Buca2012,groot2022,ma2024}. From this definition, if the state $\rho$ has a strong symmetry, its corresponding weak symmetry is satisfied.

The notion of the strong and weak symmetries is further defined 
for a general quantum channel described in Eq.~(\ref{Kraus_sum}) in Appendix A. 
Here, the strong symmetry on the channel is defined as 
$$
\hat{E}_{\ell}U_{Z_2}=e^{i\theta} U_{Z_2} \hat{E}_{\ell},
$$ for all $\ell$. 
On the other hand, the weak symmetry is defined as 
$$
U_{Z_2}\biggl[\sum_{\ell}\hat{E}_{\ell} \rho \hat{E}^\dagger_{\ell}\biggr]U^\dagger_{Z_2}=\hat{\mathcal{E}}(\rho).
$$

This condition does not require that each Kraus operator commutes with non-trivial generator $U_{Z_2}$. 

\section{Parent-Hamiltonian construction}
We review the construction of a parent Hamiltonian for a state obtained by applying a doubled-space filter $\hat{\mathcal E}$ \cite{Lee2025}. Let $H^{d,(0)}$ be a positive-semidefinite Hamiltonian with zero-energy ground state $|\phi_0\rangle\rangle$, written as
\begin{eqnarray}
H^{d,(0)}
&=&\sum_{j,\alpha=u,\ell}\hat{A}^{\dagger}_{j,\alpha}\hat{A}_{j,\alpha},
\end{eqnarray}
where $\hat{A}_{j,\alpha}|\phi_0\rangle\rangle=0$. For the filtered state $|\rho^D\rangle\rangle\equiv\hat{\mathcal E}|\phi_0\rangle\rangle$, the transformed annihilation operators are
$\hat{A}^D_{j,\alpha}=\hat{\mathcal E}\hat{A}_{j,\alpha}\hat{\mathcal E}^{-1}$ \cite{SciPostPhysCore.4.4.027,Tantivasadakarn2023,Lee2025},
which satisfy $\hat{A}^D_{j,\alpha}|\rho^D\rangle\rangle=0$. An exact parent Hamiltonian is therefore
$H^{d,(D)}
=\sum_{j,\alpha=u,\ell}\hat{A}^{D\dagger}_{j,\alpha}\hat{A}^D_{j,\alpha}$.
This construction yields the ASPT parent Hamiltonian in Eq.~(\ref{Heff}) \cite{Lee2025} and the exact parent Hamiltonians used in the main text. At the noninvertible projector limits, the resulting states are instead characterized directly by the projector constraints.

For completeness, the exact parent Hamiltonian in Eq.~(\ref{HXeD_eff}) is
\begin{eqnarray}
H^{X_e,D}_{\rm eff}
=
\sum_{m}\frac{1}{4}\biggl[
&&1+\cosh^2(4\lambda_e)
-2\cosh^2(2\lambda_e)C^\eta_{2m+1}\nonumber\\
&&-2\sinh^2(2\lambda_e)C^\eta_{2m+1}\eta^x_{2m}\eta^x_{2m+2}\nonumber\\
&&-\cosh(4\lambda_e)\sinh(4\lambda_e)
\left(\eta^x_{2m}+\eta^x_{2m+2}\right)\nonumber\\
&&+\sinh^2(4\lambda_e)\eta^x_{2m}\eta^x_{2m+2}
\biggr].
\end{eqnarray}
For a small $\lambda_e$ expansion, the form corresponds to the Hamiltonian $H_{\rm ee}$ of Eq.~(\ref{H_ee}).

\section{Mean-field Hamiltonian from $H^{\rm ps}_{\rm ee}$}
We derive the mean-field Hamiltonian associated with the qualitative effective Hamiltonian in Eq.~(\ref{H_ps}). In the conserved sector $C^\tau_{2m+1}=C^\eta_{2m+1}=+1$, Eq.~(\ref{H_ps}) reduces to
\begin{eqnarray}
H_{\rm ee}^{\rm ps}
&=&-\frac{\cosh 2\beta}{2}\sum_m C^\tau_{2m}\left(1+C^\eta_{2m}\right)
-\frac{\sinh 4\beta}{2}\sum_m\tau^z_{2m}\nonumber\\
&&-h_{\rm ps}\sum_m\eta^x_{2m+1}(1+\tau^x_{2m+1})+({\rm const.}).
\end{eqnarray}
Using
$\tau^x_{2m+1}=\tau^z_{2m}\tau^z_{2m+2}$ and $\eta^x_{2m+1}=\eta^z_{2m}\eta^z_{2m+2}$, we obtain
\begin{eqnarray}
H_{\rm ee}^{\rm ps}
&=&-\frac{\cosh 2\beta}{2}\sum_m C^\tau_{2m}(1+C^\eta_{2m})
-\frac{\sinh 4\beta}{2}\sum_m\tau^z_{2m}\nonumber\\
&&-h_{\rm ps}\sum_m\eta^z_{2m}\eta^z_{2m+2}
\left(1+\tau^z_{2m}\tau^z_{2m+2}\right)+({\rm const.}).\nonumber\\
\end{eqnarray}
We then apply the mean-field approximation \cite{wen2004,Fradkin2013},
$
C^\tau_{2m}C^\eta_{2m}
\simeq
\langle C^\eta_{2m}\rangle C^\tau_{2m}
+\langle C^\tau_{2m}\rangle C^\eta_{2m}
-\langle C^\tau_{2m}\rangle\langle C^\eta_{2m}\rangle$ 
and $\tau^z_{2m}\tau^z_{2m+2}\eta^z_{2m}\eta^z_{2m+2}\simeq
\langle\eta^z_{2m}\eta^z_{2m+2}\rangle\tau^z_{2m}\tau^z_{2m+2}
+\langle\tau^z_{2m}\tau^z_{2m+2}\rangle\eta^z_{2m}\eta^z_{2m+2}
-\langle\tau^z_{2m}\tau^z_{2m+2}\rangle
\langle\eta^z_{2m}\eta^z_{2m+2}\rangle$.
The resulting mean-field Hamiltonian is $H^{\rm ps}_{\rm MF}\equiv H^{\rm MF}_{\tau}+H^{\rm MF}_{\eta}$, whose explicit form is given in Eqs.~(\ref{MFH_1}) and (\ref{MFH_2}).\\

\section{Operator space entanglement entropy}
We briefly discuss the OSEE.
We consider a density matrix $\rho$ and its normalized Choi state $|\tilde{\rho}\rangle\rangle$. 
The density matrix for the Choi state is defined as
$
\tilde{\rho}\equiv|\tilde{\rho}\rangle\rangle \langle\langle\tilde{\rho}|
$.
We partition the total system into two subsystems, $A$ and $B$, and introduce the same spatial partition for the upper and lower bases as $|i\rangle_{u}\to |(i_A,i_B)\rangle_u$ and $|j\rangle_{\ell}\to |(j_A,j_B)\rangle_{\ell}$. 
For this partition, the reduced density matrix of subsystem $A$ is
\begin{eqnarray}
&&\tilde{\rho}_A=\Tr_B[\tilde{\rho}]\nonumber\\
&&=\sum_{(i_A,i'_A),(j_A,j'_A)}\tilde{\rho}'_{A,(i_A,j_A),(i'_A,j'_A)}
[|i_A\rangle_u|j_A\rangle_\ell {}_{u}\langle i'_A|{}_{\ell}\langle j'_A|],\nonumber\\
\\
&&\tilde{\rho}'_{A,(i_A,j_A),(i'_A,j'_A)}\nonumber\\
&&\equiv\sum_{i_B,j_B}\frac{1}{\Tr[\rho^2]}
\rho_{(j_A,j_B),(i_A,i_B)}
\rho^{*}_{(j'_A,j_B),(i'_A,i_B)}.
\end{eqnarray}
The form of the reduced density matrix $\tilde{\rho}'_{A,(i_A,j_A),(i'_A,j'_A)}$ corresponds to that used in the OSEE \cite{Prosen2007}. That is, in the formulation of the OSEE, we take the density matrix $\rho$ as the target operator. Similar choices of the target operator have also been considered in Refs.~\cite{Znidaric2008,Nieuwenburg2014}. 
Based on this correspondence, the entanglement entropy of subsystem $A$ for the state $|\tilde{\rho}\rangle\rangle$ is
$
S^{\rm OS}(A)=-\Tr[\tilde{\rho}'_{A}\ln \tilde{\rho}'_{A}],
$
which corresponds to the OSEE of the density matrix $\rho$. This quantity is used as a correlation measure between subsystems $A$ and $B$. 
The OSEE has been widely used to characterize the correlation structure and complexity of mixed-state density operators.
In particular, its system-size scaling has been shown to provide a useful diagnostic of nonequilibrium phase transitions in open quantum steady states
\cite{Prosen_PhysRevLett,Pizorn2009}, while more recent studies have employed it to characterize the buildup and suppression of correlations under dissipative dynamics \cite{Wellnitz2022}.

\nocite{apsrev42Control}
\bibliographystyle{apsrev4-2}
\bibliography{ref_v2}

\end{document}